\documentclass[pra,preprint,showpacs]{revtex4}

\usepackage{graphicx}
\usepackage{amsbsy}

\begin{document}


\title{Nondestructive interaction-free atom-photon controlled-NOT gate}
\author{Mladen Pavi\v{c}i\'{c}}
\email{pavicic@grad.hr}
\homepage{http://m3k.grad.hr/pavicic}
\affiliation{University of Zagreb, Zagreb, Croatia}

\date{\today}

\begin{abstract}

We present a probabilistic (ideally 50\%) nondestructive 
interaction-free atom-photon controlled-NOT gate, where {\em 
nondestructive} means that all four outgoing target photon 
modes of the gate are available and feed-forwardable. 
Individual atoms are controlled by a stimulated Raman adiabatic 
passage transition and photons by a ring resonator with two 
outgoing ports. Realistic estimates we obtain for ions 
confined in a Paul trap around which the resonator is mounted 
show that a strong atom-photon coupling can be achieved. 
It is also shown how the resonator can be used for controlling 
superposition of atom states. 
\end{abstract}


\maketitle

\section{\label{sec:intro}INTRODUCTION}

Individual qubits with the help of which we attempt to 
build our would-be quantum computers are fragile because 
we must have full control over the states of each of them 
separately and of all of them collectively within any time window. 
In many proposals qubits are held in traps---electrical, magnetic, 
or optical. Controlling their states in traps over time and 
getting information from them during calculations can    
decohere and unstabilize them. The less energy 
transferred by probes, usually flying qubits, to a qubit 
sitting in a trap, the better. It would be the best to have no 
energy transfer at all. Quantum mechanics provides a way to 
do this. Whenever we {\em fail} to detect a quantum system in 
one of the paths it would take to interfere with itself, we 
``erase'' the interference  fringes we would have got, if we 
had not carried out any detection or 
measurement at all \cite{renninger,dicke,my-PhD-int-f-c-osa}.
Dicke called such a measurement an {\em interaction-free 
measurement\/} \cite{dicke} and Elitzur and Vaidman realized that such  
measurements might be useful whenever we do not want to 
transfer energy to a measured object \cite{elitzur-vaidman}.
The measurements are also called 
{\em energy-exchange-free\/} \cite{pav-pla-96}, 
{\em absorption-free\/} \cite{mitch-mass-01}, and 
{\em counterfactual\/} \cite{mitch-jozsa-01} measurements, as well as 
{\em quantum interrogations\/} \cite{gilc02}. 

Two implementations of interaction-free measurements in 
quantum computation have recently been proposed. One, proposed 
by Richard Jozsa \cite{jozsa-99}, considers atoms as quantum 
computers (or parts of them) and flying qubits 
(photons) as their switches. Thus one of the two possible 
states of an atom resulting from a ``computation'' can be read 
``for free,'' with no transfer of energy to the atom, i.e.,  
without the ``computer'' actually running. Unfortunately, it 
turns out that the other state cannot be obtained for 
free \cite{mitch-jozsa-01}. 
The controlled-NOT (CNOT) gate used in this approach has 
photon states as its control qubits and output register 
states as its target qubits. 

The other proposal considers interaction-free measurements
that are used to implement essential parts of 
quantum computations, notably CNOT gates and 
entanglement \cite{azuma03,azuma04,gilc02,horod01,methot01}.
CNOT gates are essential for quantum computing because they 
enable us to set up any quantum gate and therefore to implement 
any available algorithm. Entanglement appears in almost 
all blueprints of quantum computer candidates. 

In constructing an interaction-free CNOT gate, Hiroo 
Azuma \cite{azuma03} uses a positron as a straight moving 
control qubit that blocks ($N$ times, where $N\to\infty$) 
the paths of a zig-zagging electron---a 
target qubit. The CNOT gate is constructed not directly 
but using Bell-basis measurements by the method of
Gottesman-Chuang \cite{gottesman-99}. 
In the M\'{e}thot-Wicker method, the qubits of the 
CNOT gate are two-level atoms, and photons only pass 
information between atoms \cite{methot01}. 
Gilchrist, White, and Munro \cite{gilc02} do use atoms as 
control qubits and photons as target qubits in a quantum 
Zeno setup, but to obtain a gate which could be called a 
{\em destructive} CNOT gate, 
since one of the target qubits must be destroyed and 
therefore one of the four outputs of the standard CNOT 
gate is not available. 

In this paper we propose a {\em nondestructive} CNOT gate in 
which a two-level atom in a trap is the control qubit and a 
photon interrogating it---without transferring a single 
quantum of energy to it---is the target qubit. 
``Nondestructive'' means that all four modes of the gate 
are available.\ \cite{zhao-cnot05} The proposal is an 
elaboration of the CNOT-gate setup put forward in 
Ref.\ \cite[p.\ 166]{pavicic-book-05}. 
A particular feature of the gate is that the target qubit can also 
{\it physically control\/} its control qubit by preventing the 
latter from entering into a superposition of its two 
available states. We carry out this {\em interaction-free} 
interrogation with the help of a photon resonator, and we 
prepare the atom by {\em stimulated Raman adiabatic 
passage}, STIRAP. 

We organize the paper as follows. In Sections
\ref{sec:resonator} and \ref{sec:stirap} 
we briefly present those details of interaction-free and 
STIRAP experiments (respectively) that are indispensable 
for understanding the construction of our CNOT gate. In 
Sec.~\ref{sec:i-f-cnot} we present the interaction-free 
CNOT gate itself, and in  Sec.~\ref{sec:suppr-sup} an 
interaction-free control of superposition. The conclusion 
of the paper is given in Sec.~\ref{sec:concl}. 

\section{\label{sec:resonator}THE RESONATOR}
\begin{figure}
\includegraphics[width=0.99\textwidth]{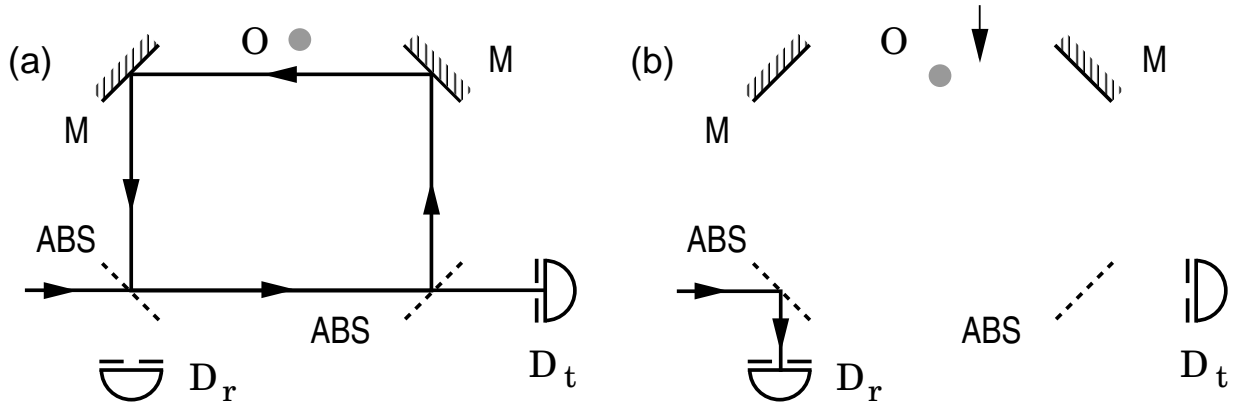}
\caption{Schematic of an interaction-free device 
according to Ref.~\cite{pav-evanston}. A single photon 
enters the resonator. M's are perfect mirrors 
(total-reflection Pellin--Broca prisms can be substituted 
for M's for higher efficiency \cite{pav-evanston}). ABS's 
are highly asymmetrical mirrors with $R=0,999$ or higher 
(with frustrated total reflection, 
i.e., optical tunnelling \cite{pav-evanston}); 
(a) When there is no object in its path
the photon exits into \it D\rm$_t$ (with a realistic 
efficiency of over 98\%); (b) When there is an object in 
its path, the photon is reflected into \it D\rm$_r$.}
\label{fig:pellin}
\end{figure}

Let us consider the setup proposed by Paul and 
Pavi\v{c}i\'{c} in 1996 \cite{ppijtp96,pav-pla-96,p-p-josab97} 
and shown in Fig.~\ref{fig:pellin}. The predictions were
confirmed in an actual experiment carried out by Tsegaye, Goobar, 
Karlsson, Bj\"{o}rk, Loh, and Lim in 1998 \cite{bjork-karlsson-98}.

We make use of a resonator which consists of two perfect mirrors 
and two highly asymmetrical mirrors that determine photon 
round trips as shown in Fig.~\ref{fig:pellin}. Other setups that 
minimize reflection losses are also possible 
\cite{ppijtp96,p-p-josab97}. A laser beam enters the resonator 
through highly asymmetrical beam splitter ABS.  
When there is no object in the resonator, an incoming laser
beam is almost completely transmitted into detector \it D\rm$_t$. 
When there is an object, the beam is almost totally 
reflected into detector \it D\rm$_r$. 
To increase efficiency, frustrated 
total reflection (which is an optical version of quantum 
mechanical tunnelling) can be used 
instead \cite{ppijtp96,p-p-josab97,pav-evanston}.
Thus the reflectivity $R$ can reach  0.9999 or higher.  
The uniqueness of the reflectivity at the beam splitters and 
perfect mirrors is assured by choosing the orientation of the 
polarization of the incoming laser beam perpendicular to the 
plane of incidence. The source of the incoming beam should be a
continuous wave (cw) laser (e.g., Nd:YAG), because
of its coherence length (up to 300$\>$km) and because of its 
very narrow linewidth (down to 10$\>$kHz in the visible range).

Each subsequent round trip contributes to a geometric 
progression whose infinite sum in the plane wave approach 
yields the total amplitude of the 
reflected beam:    
\begin{eqnarray}
B=-A\sqrt{R}{1-e^{i\psi}\over1-R\,e^{i\psi}}\,,
\label{eq:total}
\end{eqnarray}
where $\psi=(\omega-\omega_{res})T$ is the phase added by the
round trip, $\omega$ is the frequency of the incoming beam, 
$T$ is the round-trip time, and $\omega_{res}$ is the resonance  
frequency corresponding to a wavelength which satisfies 
$\lambda/2=L/j$, where $L$ is the round-trip length of the cavity 
and $j$ is an integer. 
We see that, in the long run, for any $R<1$ 
and $\omega=\omega_{res}$ we get no reflection at all---i.e., no 
response from $D_r$---if nothing obstructs the 
round trip, and almost a perfect reflection when the object blocks the 
round trip and $R$ is close to one. In terms of single photons 
(obtained by attenuating the intensity of the laser until 
the chance of having more than one photon at a time becomes 
negligible) the probability of detector $D_r$ reacting when 
there is no object in the system is zero. A response from $D_r$ indicates 
an interaction-free detection of an object in the system. The 
probability of the response is $R$, the probability of making the 
object absorb the photon $R(1-R)$, and the probability of a photon 
exiting into $D_t$ detector $(1-R)^2$.

Detailed wave packet calculations based on classical optical 
interference (analogous to the calculations for laser resonators) 
are carried out in Refs.~\cite{p-p-josab97,ppfphy98,pav-evanston} 
and they yield the following efficiencies of the suppression 
of the reflection ($r$) into $D_r$ and of the throughput ($t$)
into $D_t$ when there is no object in the resonator:
\begin{eqnarray}
r=(1-R)(1-\rho^2 R)\,\Phi, \qquad\qquad
t=(1-R)^2\,\Phi
\,,\label{eq:tau}
\end{eqnarray}
where $\rho\le1$ is a measure of overall losses and 
\begin{eqnarray}
\Phi=
{\displaystyle\int_0^\infty{\displaystyle\exp[-
{\cal T}^2(\omega-
\omega_{res})^2/2]d\omega\over\displaystyle1-
2\rho R \cos[(\omega-
\omega_{res})T]+\rho^2R^2}\over\displaystyle\int_0^\infty
\exp[-{\cal T}^2(\omega-\omega_{res})^2]d\omega}
\,,\label{eq:phi}
\end{eqnarray}
where ${\cal T}$ is the coherence time and $T$
the round-trip time.

\begin{figure}
\includegraphics[width=0.49\textwidth]{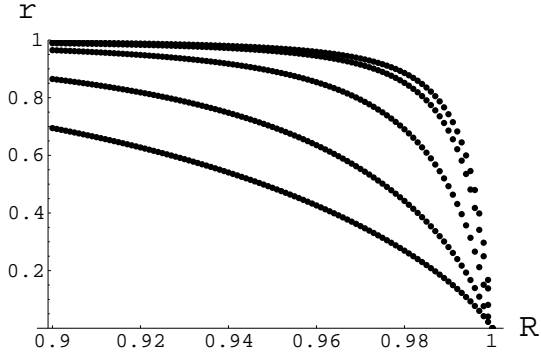}
\caption{$r$  as a function of ${\cal T}/T$ 
for $\rho=0.99$ and 
$0.9\le R \le 1$:  
${\cal T}/T=500$ (top), 150, 50, 20, and 10 (bottom). 
The differences in the shapes stem from the amount of 
losses.}
\label{fig:round-trip}
\end{figure}

In effect, the resonator has to be ``charged'' to yield a 
superposition, i.e., we have to allow the beam a sufficient 
number of round-trips to build up a destructive or 
constructive interference even when it contains just one photon. 
This corresponds to the sum which we used to obtain 
Eq.~(\ref{eq:total}) and it is shown in Fig.~\ref{fig:round-trip}.
  
Now the difference between the classical and the quantum 
picture of interference lies in the statistical behaviour
of the flying quantum system---the photon. The classical 
approach does not permit an interaction-free detection, because 
there is always an exchange of energy, e.g.~$\hbar\omega/100$. 
In the quantum approach, there can be no exchange of energy 
smaller than a quantum of energy $\hbar\omega$ corresponding to a 
single photon, and therefore only in the long run and on average 
the quantum energy transferred to an object does equal the 
classical energy. 

Therefore we cannot narrow down the time window so as to make 
the coherence time less than the time required for 
interference to build up (at least 100 round trips).  
If we did so, a photon could not enter the resonator whether or 
not the object was in the photon path. As a consequence, 
downconverted photons (the signal to enter the resonator 
and the idler to control the event) are not suitable sources 
of photons, because their coherence time is too short (in the 
range of picoseconds). The use of a cw laser or a low emission 
LED and their inability to control the number of photons within 
the time window do not, however, pose a problem to our interaction-free 
CNOT gate, because we have two distinct outgoing ports, and 
because the time required for a sufficient number of round-trips 
is a few nanoseconds, which is short enough for use  
in quantum computation, where the decoherence typically ranges 
between nanoseconds and seconds. 

\section{\label{sec:stirap}DARK STATES AND SUPERPOSITIONS}

The purpose of interaction-free detection of a macroscopic 
object is to wipe out photon interference fringes so as 
to put the object in a photon path. In the case of atoms we 
do not physically block the photon path but make them opaque 
or transparent by bringing them into states in which 
they can or cannot absorb a photon of a chosen frequency. 
This also means that if an atom can be in a superposition 
of such two states, its interaction-free interrogation by 
a photon will prevent it from entering the superposition.  
In this section we present a setup which can be used to 
make an atom (in)visible to a photon in a resonator and 
to build up a CNOT gate.

Let us consider the rubidium isotope $^{87}$Rb \cite{yu-pra04}.
(We can use many other atoms and ions that enable $\Lambda$ 
scheme presented below; e.g., $^{40}$Ca$^+$ that we discuss in 
the next section.)  
It has the closed shells $nl=1s, 2s, 2p, 3s, 3p, 3d, 4s$, $4p$, 
and one electron in the $5s$ shell, which is pushed below the $4d$ 
and $4f$ shells by spin--orbit interaction. Thus $^{87}$Rb
behaves like a system with one electron in the $5s$ ground 
state. The total angular momentum is given by {\bf J=L+S}. 
For the ground state $5s$, 
we have $s=1/2$ and $l=0$ and therefore $j=1/2$. The first 
excited states are the $5p$ states $5p_{1/2}$ and 
$5p_{3/2}$, corresponding to $s=1/2$, $l=1$, $j=3/2$, and 
$j=5/2$, respectively. They are separated by the 
spin--orbit interaction ${\mathbf L}\cdot{\mathbf S}$. 
We will consider only $j=3/2$. 

\begin{figure}
\includegraphics[width=0.99\textwidth]{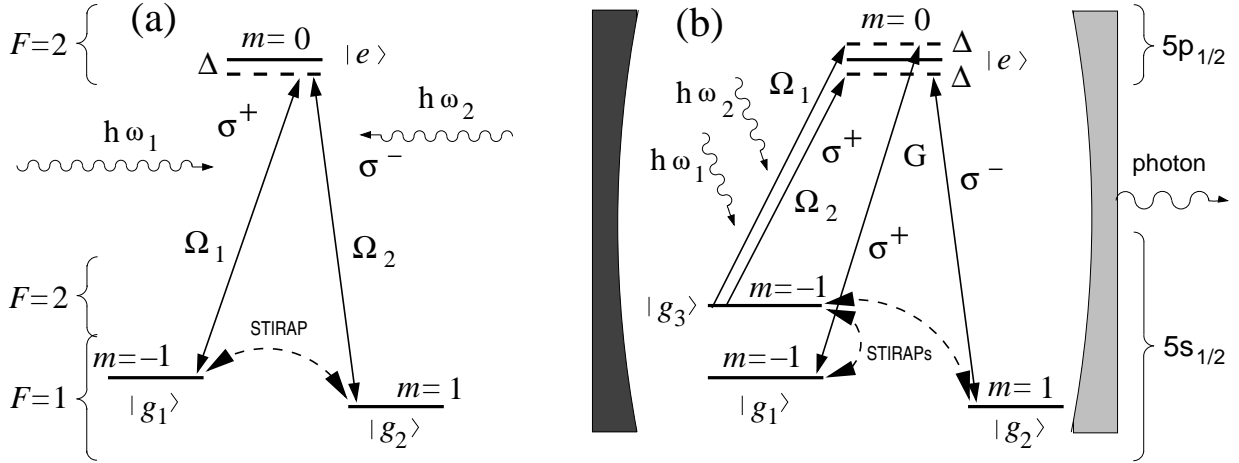}
\caption{(a) STIRAP $|g_1\rangle\leftrightarrow|g_2\rangle$
in which the population of $|e\rangle$ is completely
avoided is obtained by subsequent application of
two laser beams of frequencies $\omega_2$ and $\omega_1$,
detuned by amount $\Delta$. The $\omega_2$ one is called
the {\em Stokes} beam and it corresponds to
Rabi frequency $\Omega_2$. The $\omega_1$ one is called
the {\em pump} beam (with Rabi frequency $\Omega_1$);
(b) Two pump beams ($\Omega_1$,
$\Omega_2$) and a cavity with atom--cavity coupling ($G$)
instead of the Stokes laser beams produce superposition
 $\alpha|g_1\rangle +\beta|g_2\rangle$.}
\label{fig:rubidium}
\end{figure}

The total nuclear angular momentum {\bf K} combines with 
{\bf J} to give the total angular momentum of the atom: 
${\mathbf F}={\mathbf J}+{\mathbf K}$. 
$^{87}$Rb has $K=3/2$, and its $j=1/2$ 
ground states are split by hyperfine interaction 
into doublets with $F=K\pm j=3/2\pm 1/2=2,1$. 
Now we apply an external 
magnetic field {\bf B} to the atom to split the levels 
into magnetic Zeeman sublevels 
with magnetic quantum numbers $m=-F,-F+1,\dots,F$. 
The levels are given in Fig.~\ref{fig:rubidium} 
(cf.~Ref.~\cite{87rb}).

To excite and deexcite electrons between 
$m=\pm 1$ and $m=0$ we must use circularly polarized
photons with angular momentum $j_p=1$ and 
two additional degrees of freedom (eigenvalues of 
${\mathbf k}\cdot{\bf j}_p/k$) denoted $m_{j_p}=\pm 1$ 
\cite{messiah}. Linearly polarized photons 
cannot be used because the selection 
rules require $\Delta m=0$ for them.

When an atom absorbs a circularly polarized photon, it 
absorbs its energy and receives its angular 
momentum in its transition from the ground state to 
the excited state, and therefore the following selection 
rules must be met:  
\begin{eqnarray}
\Delta l=\pm1, \qquad 
\Delta m=m_{j_p}=\pm 1.\quad
\label{eq:select-rules}
\end{eqnarray}
When a photon is emitted, the same selection rules must 
be observed. Thus for $\Delta m=\pm 1$ we get a circularly
polarized photon and for $\Delta m=0$ a linearly 
polarized photon.

By solving the Schr\"{o}dinger equation for our three-level system
\begin{eqnarray}
\hat H|\Psi\rangle=i\hbar\frac{\partial|\Psi\rangle}{\partial t}\,,
\label{eq:sch}
\end{eqnarray} 
we arrive (after starting with a more general Hamiltonian, doing some 
approximations, and re-introducing an intermediate form of 
the wave function) at the following Hamiltonian
\begin{eqnarray}
\hat H=\frac{\hbar}{2}\left[ \begin{array}{ccc}
         0 & \Omega_1(t) & 0 \\
\Omega_1(t) &  2\Delta & \Omega_2(t)  \\ 
         0          &  \Omega_2(t)   & 0 \\ 
       \end{array}
   \right],\quad
\label{eq:3lev-Ham}
\end{eqnarray}
where Rabi frequencies $\Omega_1$ and $\Omega_2$ 
(coefficients of the general solution to 
Eq.~(\ref{eq:sch})) 
correspond to two pump laser 
beams of frequencies $\omega_1$ and $\omega_2$ that are 
detuned from resonance for $\Delta=\omega_{eg_1}-\omega_1=
\omega_{eg_2}-\omega_2$. Hamiltonian (\ref{eq:3lev-Ham})
has three eigenstates that are linear combinations of 
$|g_1\rangle$, $|g_2\rangle$, and 
$|e\rangle$. One of them is \cite{kuk-hioe-bergm-89}: 
\begin{eqnarray}
|\Psi^0\rangle=\frac{1}{\sqrt{\Omega_1^2(t)+\Omega_2^2(t)}}
\Big(\Omega_2(t)|g_1\rangle-\Omega_1(t)|g_2\rangle\Big)\,.
\label{eq:psi-0}
\end{eqnarray}
We see that this state is completely independent of the 
intermediate state $|e\rangle$ and that its 
eigenvalue---being zero---is independent of the Rabi 
frequencies $\Omega_1$ and $\Omega_2$. We call states 
$|g_1\rangle$ and $|g_2\rangle$ {\em dark states}.
Experimentally, we would obtain complete population 
transfer: 
 \begin{eqnarray}
\left|\langle g_1|\Psi^0\rangle\right|^2=1\quad{\rm for}\quad 
t\to -\infty,\qquad\ 
\left|\langle g_2|\Psi^0\rangle\right|^2=1\quad{\rm for}\quad 
t\to +\infty,\quad
\label{eq:popul-transfer-prob}
\end{eqnarray}
if we assumed $\left.\frac{\Omega_1(t)}{\Omega_2(t)}
\right|_{t\to -\infty}\to 0 \ {\rm and} \ 
\left.\frac{\Omega_2(t)}{\Omega_1(t)}\right|_{t\to +\infty}\to 0$, 
and this corresponds to switching on and off the 
second laser before switching on and off the first one. When the   
transfer $|g_1\rangle\to|g_2\rangle$ is {\em adiabatic} 
(the laser beams are gradually switched on and off; for the 
adiabaticity criteria see Ref.~\cite{berg-rmp98}), 
the system prepared in $|\Psi^0\rangle$ remains in this state 
at all times  and the process is called STIRAP (Stimulated Raman 
Adiabatic Passage) \cite{berg-rmp98}. The first laser beam 
(historically called the {\em pump beam}) is right-hand circularly 
polarized, denoted $\sigma^+$  
(because the transition $|g_1\rangle\to|e\rangle$ requires it) 
and the second beam (historically called the {\em Stokes beam})
is left-hand circularly polarized, 
denoted $\sigma^-$ (for $|e\rangle\to|g_2\rangle$).

In a quantum computer, control of single states $|g_1\rangle$ 
and $|g_2\rangle$ is less important than control of their 
superposition $\alpha|g_1\rangle+\beta|g_2\rangle$. We can 
obtain a superposition by carrying out two simultaneous 
STIRAPs to $|g_1\rangle$ and $|g_2\rangle$ from a common 
third one $|g_3\rangle$ as shown in Fig.~\ref{fig:rubidium}. 
Many such designs for controlling and transferring
superpositions have been proposed and implemented  
recently \cite{pellizzari-97,bose-vedral-prl99,yu-pra04}.
Cavities are often used instead of the second (Stokes) laser 
beams in each STIRAP \cite{yu-pra04} and we consider 
such a design. 

In Fig.~\ref{fig:rubidium} (b) a schematic is given of a 
strongly coupled atom-cavity system where the cavity is 
tuned (by shifting the mirrors) to the same frequency the 
Stokes beam would have for each transition. The cavity stimulates 
the population of levels $g_1$ and $g_2$ in the same way the 
Stokes laser fields would, and therefore the whole process 
is characterized by the following Hamiltonian: 
\begin{eqnarray}
\hat H=\frac{\hbar}{2}\left[ \begin{array}{ccc}
         0 & \Omega_i & 0 \\
\Omega_i &  2\Delta & 2G  \\ 
         0          & 2G   & 0 \\ 
       \end{array}
   \right],\quad
\label{eq:cav-Ham}
\end{eqnarray}
where $i=1,2$ and 
$G=\sqrt{\hbar\omega/(2\varepsilon_0 V_{\rm cavity})}$ 
is the {\em atom-cavity coupling constant} ($V_{\rm cavity}$ 
is the cavity mode volume). The photon which supports 
the cavity modes and the population of  $g_1$ and $g_1$ 
levels eventually leaks from the cavity. 

The state of the atom coupled to the cavity photon state is: 
\begin{eqnarray} 
|\Psi(t)\rangle&=&\frac{\alpha}{\sqrt{4G^2+\Omega_1(t)}}
(2G|g_3,\emptyset\rangle-\Omega_1(t)|g_1,R\rangle)\nonumber\\ 
&&+\ \ \frac{\beta}{\sqrt{4G^2+\Omega_2(t)}}
(2G|g_3,\emptyset\rangle-\Omega_2(t)|g_2,L\rangle).\qquad\qquad
\label{eq:two-dark-states-b}
\end{eqnarray} 
Thus at the beginning of the STIRAP process, the system 
is in state $|g_3,\emptyset\rangle$,  where $|\emptyset\rangle$ 
means that there is no cavity photon coupled to $|g_s\rangle$. 
As $\Omega_1$ and $\Omega_2$ gradually increase, the system 
adiabatically evolves to state
\begin{eqnarray} 
|\Psi(t)\rangle=\alpha|g_1,R\rangle+
\beta|g_2,L\rangle\,,
\label{eq:stirap-state-e-p}
\end{eqnarray}
and when the photon in the state $|R\rangle+|L\rangle$
leaves the cavity, the atom state {\em jumps} 
\cite{plenio-knight-rmp98} into the required superposition:
\begin{eqnarray} 
|\Psi\rangle=\alpha|g_1\rangle+\beta|g_2\rangle\,.
\label{eq:stirap-state-e}
\end{eqnarray}
The jump is probabilistic and has a success probability 
of 50\%.

\section{\label{sec:i-f-cnot}INTERACTION-FREE CNOT GATE}

In this section we show how the resonator described in 
Sec.~\ref{sec:resonator} can be used to construct an 
interaction-free CNOT gate and then in Sec.~\ref{sec:suppr-sup}
we discuss how to control and suppress an atom superposition  
obtained during quantum computation. 

To construct an interaction-free CNOT gate, we substitute an 
atom, for example $^{87}$Rb of Sec.~\ref{sec:stirap}, 
for the object in our resonator in Fig.~\ref{fig:pellin}. 
The $^{87}$Rb atom will be transparent to properly 
polarized photons of specific frequency when there is 
no electron in the ground level that a photon 
could excite to a higher level (a photon will not ``see''
the atom), and nontransparent when there is an electron 
populating the ground level. 

In Sec.~\ref{sec:stirap} we saw that a left-hand 
circularly polarized photon can excite an atom from 
its ground state $|g_1\rangle$ ($5s_{1/2},\ F=1,\ m=-1$) to 
its excited state $|e\rangle$  ($5p_{1/2},\ F=2,\ m=0$), and that 
the right-hand circularly polarized photon can excite the atom 
from $|g_2\rangle$ ($5s_{1/2},\ F=1,\ m=+1$) to $|e\rangle$ 
($5p_{1/2},\ F=2,\ m=0$). So an $L$-photon will ``see'' 
the atom in $|g_1\rangle$ but will not ``see'' it when it is 
in  $|g_2\rangle$. With an $R$-photon, the opposite is true. 
The energy differences to the detuned excited level 
are the same, so both photons have the same frequency 
(as the ``G part'' of Fig.\ \ref{fig:rubidium}$\>$(b): 
$|g_1\rangle\to\Delta$ and $|g_2\rangle\to\Delta$). 
We can induce a change of the atom from $|g_1\rangle$ to 
$|g_2\rangle$ and back by a STIRAP process, with two additional 
external laser beams, as explained in the previous section. 

To build our CNOT gate we use the resonator we introduced  
in Sec.\ \ref{sec:resonator}. When a photon does not ``see'' 
the atom, its laser beam will interfere with itself in a resonator 
so that classical and quantum descriptions of photons give the same 
result \cite{carmichael-book,mandel-wolf-book} as we already stressed
in Sec.\ \ref{sec:resonator}. On the other hand, when we say 
that a photon ``sees'' an atom, that means that the atom would 
have absorbed the photon if it had come to it---in reality it 
cannot come to it because it is being reflected from the entrance 
to the resonator (see Fig.~\ref{fig:pellin}). To show that 
the atom in $|g_1\rangle$ ($|g_1\rangle$) is realistically 
opaque for $L$ ($R$) circularly polarized photons, we must resort 
to quantum theory and we shall come back to this point in the 
second half of this section. 

This feature of our resonator approach that there are no 
photon loops in it when the atom is opaque is yet another 
advantage over a Zeno-like setup. We have to carry out quantum 
calculations only of a single absorption of a photon by the atom 
which reduces to a single strong atom-photon interaction while in 
a Zeno setup each photon loop inside a cavity with an opaque 
atom includes a strong photon-atom interaction. 
Cf.\ quantum calculations for a Zeno-like atom-photon 
interaction-free setup carried out by Luis and 
S{\'a}nchez-Soto.\ \cite{luis-98, luis-99,luis-99a}
   
We feed our resonator with $+45^\circ$ and $-45^\circ$ linearly 
polarized photons to achieve the same conditions for round trips 
of both kinds of photons within the resonator. To the right 
of the atom we place a quarter-wave plate (QWP) to turn 
a $45^\circ$-photon into an $R$-photon and a $-45^\circ$-photon 
into an $L$-photon. To the left of the atom we place a half-wave 
plate (HWP) to change the direction of the circular polarization 
and then another QWP to transform the polarization back into 
the original linear polarization. We denote the atom states as 
follows:
\begin{eqnarray} 
|0\rangle=|g_1\rangle,  \qquad |1\rangle=|g_2\rangle,
\label{eq:a-states-int-f-a}
\end{eqnarray} 
and take these atom states as control states 
and the atom itself to be our control qubit. 
We denote the photon states as follows:
\begin{eqnarray} 
|0\rangle=|45^\circ\rangle,  \qquad |1\rangle=|-45^\circ\rangle,
\label{eq:a-states-int-f-f}
\end{eqnarray} 
and we take these photon states as the target states and the photons
as target qubits. For example, $|01\rangle$ means that the atom is 
in state $|g_1\rangle$ and the photon is polarized along $-45^\circ$. 

\begin{figure}
\includegraphics[width=0.99\textwidth]{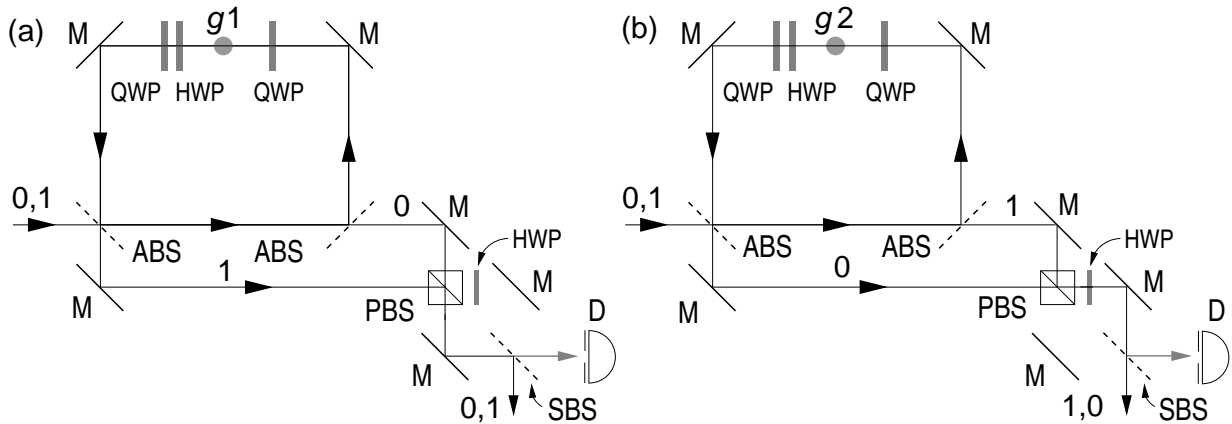}
\caption{Interaction-free CNOT. (a) The atom is in state 
$|g_1\rangle$ and can absorb a $-45^\circ$ polarized photon. 
Therefore a photon in state $|1\rangle$ cannot enter the cavity
and thus $|0\rangle\to|0\rangle$ and $|1\rangle\to|1\rangle$. 
(b) The atom is in state $|g_2\rangle$ and can absorb a 
$+45^\circ$ polarized photon. Therefore photon $|0\rangle$ 
cannot enter the cavity and thus $|0\rangle\to|1\rangle$ 
and $|1\rangle\to|0\rangle$. ABS are highly asymmetrical beam 
splitters with $R=0.999$; SBS is a symmetric 50:50 beam splitter; 
M are perfect mirrors; PBS is a polarizing beam 
splitter which lets $|0\rangle$ photons through and reflects 
$|1\rangle$ photons; HWP and QWP are half- and quarter-wave plates, 
respectively---the plates in the resonator turn linear polarization 
into circular and back into linear and HWP after PBS turns 
$|0\rangle$ ($|1\rangle$) photon into $|1\rangle$ ($|0\rangle$) 
photon;  D is a detector---ideally, when it does not click, 
the target qubit exits at the other side of SMS.} 
\label{fig:int-f-cnot} 
\end{figure}

Now consider Fig.~\ref{fig:int-f-cnot}$\>$(a). A photon in 
state $|0\rangle$ does not ``see'' the atom in state 
$|g_1\rangle$ and will therefore exit the resonator through the 
right port and will pass through the polarizing beam 
splitter PBS. A photon in state $|1\rangle$ ``sees'' the atom 
in the state $|g_1\rangle$ and therefore does not enter the resonator 
but goes down to the PBS and is reflected by it. 
Fig.~\ref{fig:int-f-cnot}$\>$(b) refers to the atom in  
state $|1\rangle$. A photon in state $|0\rangle$ sees it, 
goes down and passes through it, then goes through the 
half wave plate (HWP) which changes its state to  $|1\rangle$. 
A photon in state $|1\rangle$ does not see the atom and exits 
through the right port, reflects at PBS and changes to
state $|0\rangle$ when passing through HWP. This would 
(before the 50:50 beam splitter shown in Figure 
\ref{fig:int-f-cnot}) give us a classical reversible CNOT 
gate and a nondestructive method of detecting the 
states of an atom.

However if we wanted to integrate the obtained CNOT gate 
into the circuits of a would-be quantum computer, then we 
should make the target operation unitary, i.e., we have 
to erase the which-path information the photons carry. 
We do so with the help of a symmetrical 50:50 beam splitter
shown as SBS in Figure \ref{fig:int-f-cnot}. Ideally, when 
detector D does not response, the target qubit will exit 
at the opposite side of the beam splitter so as to yield 
the following CNOT qubit values
\begin{eqnarray} 
|00\rangle\to|00\rangle,\qquad|01\rangle\to|01\rangle,\qquad
|10\rangle\to|11\rangle,\qquad|11\rangle\to|10\rangle.\qquad
\label{eq:int-f-cnot-ver}
\end{eqnarray} 

Beam splitter SBS makes our CNOT probabilistic. Detector 
will ideally {\em not\/} response in half of the cases and this 
means that we would be able to forward on average every second 
target qubit to subsequent computation stages. Realistically, 
single photon detectors have recently reached the efficiency 
of 50\%\ \cite{s-phot-det06} and photon-number-resolving detectors 
the efficiency of 90\%\ \cite{s-phot-det05}, so, the efficiency 
of the CNOT gate would be less than 50\%. As a 
partial remedy, we could use two resonators simultaneously to 
obtain two CNOT gates consisting of one control atom qubit and 
two photon target qubits. Although statistically independent, 
their outcomes could support each other for getting more 
reliable final results, for obtaining middle stage results, 
and perhaps even for obtaining an error correction scheme for 
the target qubits (an algorithm for the purpose should be 
devised). This would enable us to compare our CNOT 
gate with the recent feed-forwardable all-optical CNOT gates.\ 
\cite{pit-frans03,pan-zeil-cnot-prl04,pit-frans05}

To estimate how efficiently an $L$-photon will ``see'' 
an atom in $|g_1\rangle$ state (and an $R$-photon an 
atom in $|g_2\rangle$ state) we have to calculate the 
probability with which an atom in $|g_1\rangle$ ($|g_2\rangle$) 
state would absorb a photon in $L$ ($R$) state. 
The total Hamiltonian for the atom-photon coupled system 
can be decomposed into three 
parts \cite{carmichael-book,mandel-wolf-book}:
\begin{eqnarray}
\hat H=\hat H_a+\hat H_p+\hat H_i,
\end{eqnarray}
where $\hat H_a$ is the Hamiltonian of the two-level atom, 
$\hat H_p$ of the photon field, and $\hat H_i$ of their 
interaction. We assume that both the atom and the field 
are quantized so as to have
\begin{eqnarray}
\hat H_i={\rm i}\omega\left(\langle g|\hat{\pmb{\mu}}|e\rangle
|g\rangle\langle e|-
\langle g|\hat{\pmb{\mu}}^*|e\rangle
|e\rangle\langle g|\right)\cdot 
\hat{\pmb{\rm{A}}}(r_0,t),
\label{eq:tot-ham}
\end{eqnarray}
where $\pmb{\mu}$ is the atomic dipole moment, 
$\hat{\bf{A}}$ is the vector potential of the electric field
of the laser beam and $\pmb{\rm r}_0$ is the position of the 
atom, which we take to be fixed. The latter assumption is 
made on the following grounds.

Within an ion trap, ionized atoms can be confined to a region 
much smaller than the optical wavelength and their position can 
be controlled with a precision of under 10 nm.\ \cite{mundt-blatt-02}    
The cavity (Sec.\ \ref{sec:stirap}) and the resonator 
can be put around the ion trap following a proposal recently 
elaborated theoretically by Maurer, Becher, Russo, Eschner, and Blatt 
\cite{blatt04} and Keller, Lange, Hayasaka, Lange, and 
Walther \cite{kell-walther} and confirmed experimentally by  
both groups \cite{mundt-blatt-02,kell-walther-apb03} for 
$^{40}$Ca$^+$. Our $^{87}$Rb cannot be easily ionized because 
$^{87}$Rb$^+$ would lack its $5s$ electron which forms our ground 
states and the first option for $^{87}$Rb$^-$ is $5p$ which builds 
our excited states. Higher $^{87}$Rb$^-$ are difficult to obtain 
and are unstable.\ \cite{rb-ion-94} But other stable ions, as e.g. 
$^{43}$Ca$^+$ have abundant available states ($^{43}$Ca$^+$ has 
nuclear spin 7/2) with a structure that enables STIRAPs of the 
kind we analysed in Sec.\ \ref{sec:stirap} for $^{87}$Rb and can 
be used instead of $^{87}$Rb$^-$. We however leave details of such 
a reformulation to a more realistic experimental future proposal 
simply because $^{87}$Rb structure and its Zeeman splitting has 
already been given a detailed  experimentally tested model.\ 
\cite{zhu-pra99,87rb} Zeeman splitting of other 
candidates, including  $^{43}$Ca$^+$, has not been sufficiently 
explored as of yet.

In the interaction picture we use the interaction Hamiltonian $H_i$ 
from Eq.~(\ref{eq:tot-ham}) to obtain the following probability for 
the photon absorption by the atom in time $\Delta t$ 
\cite{mandel-wolf-book}: 
\begin{eqnarray}
\frac{\omega_0^2}{2\hbar\omega\varepsilon_0 V}
|\langle g|\pmb{\mu}|e\rangle\cdot\pmb{\varepsilon}|^2
\cos^2\frac{1}{2}\Theta\left[\frac{\sin\frac{1}{2}
(\omega-\omega_0)\Delta t}{\frac{1}{2}(\omega-\omega_0)}\right]^2,
\label{eq:abs-prob}
\end{eqnarray}
where $\omega_0$ is the atomic frequency, $\omega$ is the 
laser field frequency, $\Theta$ is the polar angle of the atomic
Bloch vector, $\pmb{\varepsilon}$ is the polarization vector, 
and $V$ the quantization volume. 

This means that several conditions have to be satisfied 
to get a high absorption probability. First we have to tune the
laser frequency $\omega$ close enough to the atomic frequency 
$\omega_0$ and this is achieved by the level of precision already 
experimentally reached in targeting individual ions trapped in 
Paul traps within an optical cavity as we mentioned above. 
This has also been achieved experimentally in cavity quantum 
electrodynamics (CQED) by Brune, Schmidt-Kaler, Maali, Dreyer, 
Hagley, Raimond, and Haroche \cite{haroche-prl96} already ten 
years ago. They obtained a coherent exchange of photons between 
the cavity field and individual atoms (vacuum Rabi oscillations). 
On the other hand, lasers can have linewidths that are less than 
1 Hz and the precision of determining frequency of atomic 
transitions also approaches 1 Hz.\ 
\cite{hertz-precision-05} 

Next, for an atomic transition $m=0\to m=-$1 (when 
$\langle g|\pmb{\mu}|e\rangle=
|\mu|(\mathbf{x}+i\mathbf{y})/\sqrt{2}$, 
where $|\mu|=|\langle g|\pmb{\mu}|e\rangle|$) 
the probability (\ref{eq:abs-prob}) is greatest for right 
circularly polarized photons 
$\pmb{\varepsilon}=(\mathbf{x}+i\mathbf{y})/\sqrt{2}$ and 
vanishes for left circularly polarized ones 
$\pmb{\varepsilon}=(i\mathbf{x}+\mathbf{y})/\sqrt{2}$. 
Thus we have to orient the external magnetic field 
$\mathbf B$ along the direction of the beam that hits the 
atom. As for $\cos^2\frac{1}{2}\Theta$ term, $\Theta=0$ 
means that the atom is in pure $|g \rangle$ state 
and $\Theta=\pi$ that its $|g \rangle$ state is not 
populated at all.  

What remains to be evaluated to estimate the level 
of coupling of the atom to the photon field is the quantization 
volume $V$ and the terms containing $\omega$ in the probability 
(\ref{eq:abs-prob}). 
The square root of the first two terms 
\begin{eqnarray}
g=\sqrt{\frac{|\mu|^2\omega_0^2}{2\hbar\omega\varepsilon_0 V}}, 
\end{eqnarray}  
is called the {\em dipole coupling constant} \cite{carmichael-book}, 
or the {\em rate of coupling} of an atom to a single cavity mode 
\cite{kell-walther,grangier-fp00} (usually with the assumption 
that $\omega\approx\omega_0$). 
In CQED and STIRAP cavities, $g$ is compared with spontaneous 
emission from the atom (transverse damping rate $\gamma$) and with leaking 
out of the cavity (damping rate $\kappa$). The calculations for these 
cavities differ from the one carried above, but the overall results are 
comparable. For the former cavities a {\em strong coupling} is achieved 
when $g$ is much larger than both $\gamma$ and $\kappa$.\ 
\cite{carmichael-book,kell-walther,grangier-fp00} One can achieve 
this by making a cavity as small as possible 
thereby decreasing $V$ and making $g$ larger, by choosing 
atoms with a narrower atomic linewidth $\gamma$, and by decreasing
the linewidth of the cavity mode $\kappa$.  With our resonator 
cavity, however, we do not have to care about spontaneous emission 
and/or leaking from the cavity because the interaction-free 
absorption (almost) never really excites the atom and the photon 
(almost) never really reaches the atom.  
So, our resonator cavity does not have to be small. In Sec.\
\ref{sec:resonator} we have seen that  $\lambda_{\rm res}=2L/j$, 
where $L$ is the round-trip length of the cavity and $j$ is an integer. 
By picking a larger $j$ we get a larger cavity in which the 
birefringent optics would fit. 

After integrating the terms containing $\omega$ in 
(\ref{eq:abs-prob}) over $\omega$ we get the rate of absorption: 
$q^2\Delta t$ (the probability turns out to be a function of 
$(\Delta t)^2$). This rate is higher than the one we get in CQED 
experiments \cite{haroche-prl96} where the atoms move through 
a cavity with a velocity of over 100$\>$m/s and nevertheless 
couple strongly to the cavity field within less than 
100$\>\mu$s. In our resonator we can achieve much 
longer and also shorter times ($1\>{\mu\rm s}<\Delta t<1\>{\rm ms}$; 
the shortest time is limited by the resonance build-up time when a 
transparent object is in the resonator and the longest one by the 
coherence time of cw lasers). So, we can conclude that under 
realistic conditions the atom will be strongly coupled to our 
resonator cavity field. The shortest required times for the 
coupling have to be determined for a chosen atomic system.    

We should add that the Stokes and pump beams for STIRAP 
switching $|g_1\rangle\leftrightarrow|g_2\rangle$ can be used 
together with the resonator cavity inclined at a small angle to 
its beam because that beams are much stronger than the resonator 
beam and a small misalignment with the magnetic field $\mathbf B$ 
will not significantly influence the transitions.\ 
\cite{arimondo,berg-rmp98}  
We cannot carry out STIRAP transitions and run interaction-free
CNOT simultaneously because of the electromagnetically induced 
transparency (EIT) by the transitions.\ 
\cite{arimondo,lukin03,fleischhauer05} If we wanted to carry 
out interaction-free interrogations of STIRAP transitions, 
we should employ transitions that are not used by the transitions, 
e.g., $m=-1,F=1\ \to\ m=-2,F=2$ and   $m=1,F=1\ \to\ m=2,F=2$ as 
shown in Fig.\ \ref{fig:superp}. 

\section{\label{sec:suppr-sup}INTERACTION-FREE CONTROL OF 
SUPERPOSITION}

Our design can also be used to control the STIRAP 
superposition of the states discussed in Sec.\
\ref{sec:stirap}, Eq.~(\ref{eq:stirap-state-e}), in two 
ways. The first one is by determining a time span within which an 
atomic superposition can be built up and the second one is by 
suppressing the formation of atomic superposition using the 
interaction-free erasure of interference fringes obtained in 
Ref.~\cite{pav-pla-96}. 

In the STIRAP process presented in Sec.~\ref{sec:stirap} 
the electron is moved from the state $|g_3\rangle$ 
shown in Fig.~\ref{fig:rubidium}$\>$(b) into the 
superposition of the states $|g_1\rangle$ and $|g_2\rangle$
given by Eq.~(\ref{eq:stirap-state-e}). Interaction-free
detection can be used to determine, within a few microseconds, 
when the electron leaves state $|g_3\rangle$, as follows. 

\begin{figure}
\begin{center}
\includegraphics[width=0.6\textwidth]{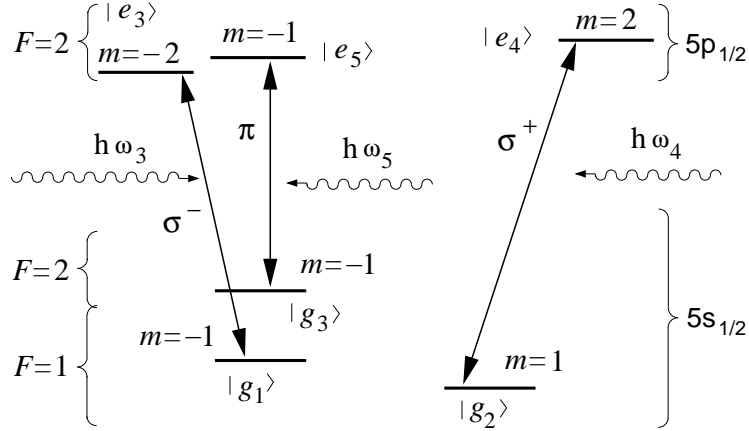}
\end{center}
\caption{Transitions which make an atom opaque for photons in 
the interaction-free resonators of Fig.~\ref{fig:int-f-cnot}. 
If an electron is in one of the $|g\rangle$ states, then a photon
will exit through one port of the gate, and if not, through 
the other.} 
\label{fig:superp}
\end{figure}

We use our interaction-free resonator with the interrogating 
paths perpendicular to the axis of the cavity and wave plates 
removed in order to probe the atom with a linearly polarized 
photon $\hbar\omega_5$, denoted $\pi$ in Fig.~\ref{fig:superp} 
(a circularly polarized photon corresponding to the transition 
$|g_3\rangle\to|e_3\rangle$ would also work). 
We tune the resonator and photons to the frequency $\omega_5$ 
that corresponds to a possible transition $|g_3\rangle\to|e_5\rangle$.
When our resonator is applied to an atom in which a STIRAP transition
$|g_3\rangle\to(\alpha|g_1\rangle+\beta|g_2\rangle)$ is induced, 
its repeated interrogation will pinpoint the time of this transition.  
This is because the photon ``sees'' the electron if 
transition $|g_3\rangle\to|e_5\rangle$ is possible and does not 
``see'' it when it leaves $|g_3\rangle$. In the former case it  
exits at one port of the resonator and in the latter at the other. 
The photon cannot of course really excite the atom, because it 
cannot transfer any energy to it.

Assuming that a realistic round trip of our resonator can be reduced 
to a few cm and calculating \cite{ppijtp96} that 200 
to 300 round trips (under 100$\>$ns) are sufficient to establish 
interference within the resonator, we calculates that 
a laser has to be {\em on} for several $\mu$s in 
order to detect whether or not an electron is in state 
$|g_3\rangle$, i.e., whether the atom is opaque or transparent. 
We must therefore reduce the intensity of a cw laser so as to 
obtain on average one photon within this time window.

The other way of controlling our STIRAP is to prevent 
building of superposition $\alpha|g_1\rangle+\beta|g_2\rangle$ and 
forcing the electron into either state $|g_1\rangle$ or 
state $|g_2\rangle$. For this purpose, we use two interaction-free 
resonators with interrogating paths oriented perpendicularly 
to the axis of the cavity and tuned to frequencies  
$\omega_3$ and $\omega_4$ with circularly polarized photons, as shown 
in Fig.~\ref{fig:superp}. Linearly polarized photons corresponding 
to transitions to $m=-1$ and $m=1$, respectively, would also work. 
Note that the scheme with photons tuned to transitions 
$|g_1\rangle\to|e_5\rangle$ and $|g_2\rangle\to|e_5\rangle$ 
would not work, because, as shown in Sec.~\ref{sec:stirap}, 
the atom is transparent to such photons in both 
cases---when there is no electron in states $|g_1\rangle$ and 
$|g_2\rangle$ and when STIRAPs are in progress. In the latter 
case we speak of electromagnetically induced transparency (EIT) 
\cite{arimondo,lukin03,fleischhauer05}.

\section{\label{sec:concl}CONCLUSION}

In conclusion, we have obtained a probabilistic interaction-free 
CNOT gate in which two atom Zeeman states represent the control 
qubit and two photon polarization states represent the target qubit. 
The gate, which is a photon ring resonator, is robust because 
it does not transfer any energy to atoms in over 95\%\ 
of tests. Unlike the previous atom-photon CNOT gate 
\cite{gilc02}, this gate has all four modes available, 
because it has two exit ports for photons, 
each of which can let photons in both polarization 
states out, depending on the state the atom is in. 

In Sec. \ref{sec:i-f-cnot} we carried out quantum calculations 
and made realistic estimations for a possible experiment. 
If we confine ions (e.g., $^{40}$Ca$^+$)  
to under 10$\>$nm by using a Paul trap and mount the 
resonator around it, we arrive at the rate of absorption 
which is higher than in other experimentally tested cavities. 
This amounts to a strong coupling between the atoms in our 
ring cavity and its field modes.     

The interaction-free resonator can also be used to 
control a stimulated Raman adiabatic passage (STIRAP) 
from an individual state to a state of superposition. 
We can control the time when an atom changes its  
state. Such detection takes under 1$\>\mu$s. 

Another control we can exert over qubits is to suppress 
atom-state superposition. The quantum system is altered
without any energy transfer to the system in observation of 
the quantum indistiguishability principle, which states that 
no information can be acquired about the population of
particular states which take part in a superposition. 

Thus the interaction-free resonator can be used in 
quantum computation to manipulate qubits and to 
obtain information on them during computation 
without decohering their states. It is suitable for 
systems that can be scaled up because it is non-destructive, 
i.e., it does not destroy the output states, and because it 
provides information on the success of the CNOT-gate 
operation that can be used for subsequent manipulation
of the same photon qubits. More specifically, we can  
amplify the null detections of detector D in Fig.\ 
\ref{fig:int-f-cnot} by combining two resonators, 
where one of them would give information on 
the output of the other. 

\begin{acknowledgments}
This work was supported by the Ministry of Science, 
Education, and Sport of Croatia, Project No.~0082222.   
\end{acknowledgments}

\end{document}